\RequirePackage{fix-cm}
\RequirePackage{rotating}
\documentclass[smallcondensed]{svjour3}     
\smartqed  
\usepackage{amstext}
\usepackage{amsmath}
\usepackage{mathtools}
\usepackage{amssymb}
\usepackage{bm}
\usepackage{mathrsfs}
\usepackage[mathscr]{eucal}
\usepackage{mathtools}
\usepackage{color}
\usepackage{graphicx}
\usepackage{graphics}
\usepackage{caption}
\usepackage{tabularx,adjustbox}
\usepackage{fullpage,rotating,booktabs,dcolumn,pdflscape}
\usepackage{cite}
\usepackage{placeins}
\usepackage{float}
\usepackage{bm}
\usepackage{amsfonts,amssymb}
\usepackage{layout}

\usepackage[  
colorlinks=true,
pdfborder={0 0 0},
citecolor=blue,
linkcolor=blue
]{hyperref}
\newcommand*{\myeqref}[2][Eq.~]{%
	\hyperref[{#2}]{#1(\ref*{#2})}%
}
\def\equationautorefname#1#2\null{%
	Eq.#1(#2\null)%
}

\makeatletter
\newcommand{\vast}{\bBigg@{4}}
\newcommand{\Vast}{\bBigg@{5}}
\makeatother

\newcommand\ddfrac[2]{{\displaystyle\frac{\displaystyle #1}{\displaystyle #2}}}
\usepackage{stmaryrd} 
\expandafter\let\csname equation*\endcsname\relax
\expandafter\let\csname endequation*\endcsname\relax
\begin{document}

\title{Soliton Solutions and Conservation Laws for a Self-interacting Scalar Field in $\Phi^{4}$ Theory}



\author{M.A.Z. Khan  \and M.Moodley \and F.Petruccione}


\institute{M.A.Z Khan \at
              Quantum Research Group, University of KwaZulu-Natal, Private Bag X54001, Durban, South Africa. \\
              \email{muhammadalzafark@gmail.com}           
           \and
           M.Moodley \at
              Quantum Research Group, University of KwaZulu-Natal, Private Bag, X54001, Durban, South Africa. \\
              \email{moodleym2@ukzn.ac.za}
              \and
           F. Petruccione \at
           Quantum Research Group, University of KwaZulu-Natal, Private Bag X54001, Durban, South Africa.
           \email{petruccione@ukzn.ac.za}
}


\maketitle

\begin{abstract}
We calculate soliton solutions to the scalar field equation of motion that arises for the $4^{\text{th}}$-order extended Lagrangian ($\Phi^{4}$ theory) in quantum field theory using the extended hyperbolic tangent and the sine-cosine methods. Using the former technique, ten complex soliton waves are obtained; we graphically represent three of these profiles using density plots. In the latter case, two real soliton solutions are obtained, of which, we demonstrate the wave profile for the positive case. Using the multiplier method, we calculate conservation laws in $\left(1+1\right)$-, $\left(2+1\right)$-, and $\left(3+1\right)$-dimensions producing three, six, and ten conservation laws respectively. Lastly, we reflect on the application of conservation laws in particle physics and phenomenology.
\end{abstract}

\section{Introduction}\label{INTRODUCTION}
Partial differential equations are the mathematical cornerstone upon which modern theoretical physics rests. They are used as applicable models in a plenitude of areas of research in physics ranging from the Lagrangian and Hamiltonian mechanics in classical mechanics, where the speeds of the objects being studied is much less than the speed of light; Maxwell's and Laplace's equations, used in the study of electrodynamic phenomena; the heat equation, used to study heat transfer; the wave equation, and its variants -- Burgers equation, the Korteweg-de Vries (KdV) equation, and the Boussinesq equation -- used to study classical wave phenomena and its quantum variants; the Schr\"{o}dinger equation, used to study quantum mechanical probabilistic phenomena; the Einstein field equations, used to study astrophysical and relativistic phenomena; and the Navier-Stokes phenomena, used to study fluid dynamics. These examples are by no means exhaustive, but they serve to underscore the enduring influence of PDEs in shaping our understanding of the universe. Nonlinear partial differential equations are particularly difficult to solve because of source and sink terms added to their linear counterparts, and the introduction of non-integrable integrals, in the form of special functions, that arise. Oftentimes, researchers resort to numerical techniques and perturbation methods. However, solutions obtained from these techniques are only valid for a limited region in space-time, and do not describe the global evolution of the system at every point. This paper is a study of the $\Phi^{4}$-equation that arises in the study of quartic interactions for a self-interacting scalar field. The fields themselves can be real or complex. For the real scalar field, we consider the Lagrangian of the form   

\begin{equation}
	\mathscr{L}(\phi;\partial_{\mu}\phi) = \frac{1}{2}(\partial_{\mu}\phi \hspace*{0.05cm} \partial^{\mu}\phi - m^{2}\phi^{2}) - \frac{1}{4!}\phi^{4}, \tag{1}\label{1}
\end{equation}

where $\phi(x,t) \in \mathbb{R}$ represents the real scalar field, and $\tilde{\lambda}$ is the dimensionless gauge parameter that is proportional to the strength of the interaction between fields. The real scalar field for the Langrangian in \myeqref[Equation~]{1} is $\textbf{Z}_{2}$ - invariant. This implies that for the transformation, $\phi \rightarrow -\phi$, we get

\begin{align}
	\widetilde{\mathscr{L}} & = \frac{1}{2}\Big[\partial_{\mu}(-\phi)\partial^{\mu}(-\phi) - m^{2}(-\phi)^{2}\Big] -\frac{\tilde{\lambda}}{4!}(-\phi)^{4} \nonumber  \\
	& = \frac{1}{2}\left(\partial_{\mu}\phi \hspace*{0.05cm} \partial^{\mu}\phi - m^{2}\phi^{2}\right) - \dfrac{1}{4!}\phi^{4} \nonumber  \\
	& = \mathscr{L}, \tag{2}\label{2}
\end{align}

and $\textbf{Z}_{2} = \{-1,+1\}$. If the square of the mass term is imaginary, i.e. $m \in \mathbb{C}$, then $ \textbf{Z}_{2}$ symmetry breaks and produces singularities. The mechanism of breaking is a result of bicameral energy minima; see \cite{ref1, ref2, ref3, ref4} for a full deliberation on symmetry breaking. 

For two scalar fields, we may extend the Lagrangian in \myeqref[Equation~]{1} as follows

\begin{equation}
	\mathscr{L}(\bar{\phi}_{1},\bar{\phi}_{2}) = \dfrac{1}{2}(\partial_{\mu}\bar{\phi}_{1}\partial^{\mu}\bar{\phi}_{1} - m^{2}\bar{\phi}_{1}^{2}) + \dfrac{1}{2}(\partial_{\mu}\bar{\phi}_{2}\partial^{\mu}\bar{\phi}_{2} - m^{2}\bar{\phi}_{2}^{2}) + \dfrac{\tilde{\lambda}}{4}(\bar{\phi}_{1}+\bar{\phi}_{2})^{2}. \tag{3.1}\label{3.1}
\end{equation}

Introducing the scalar fields, we get

\begin{align}
	\phi(x,t) = \dfrac{1}{\sqrt{2}} \left[\bar{\phi}_{1}(x,t) + i\bar{\phi}_{2}(x,t)\right], \tag{3.2.1}\label{3.2.1} \\
	\phi^{\ast}(x,t) = \dfrac{1}{\sqrt{2}} \left[\bar{\phi}_{1}(x,t) + i\bar{\phi}_{2}(x,t)\right], \tag{3.2.2}\label{3.2.2}
\end{align}

where $\phi(x,t) \in \mathbb{C}$ is the complex scalar field. It is trivial to verify that substituting  \myeqref[Equation~]{3.2.1} and \myeqref[Equation~]{3.2.2} into \myeqref[Equation~]{3.1} transforms the Lagrangian into

\begin{equation}
	\mathscr{L}(\phi,\phi^{\ast};\partial_{\mu}\phi,\partial_{\mu}\phi^{\ast}) = \partial_{\mu}\phi \hspace*{0.05cm}\partial^{\mu}\phi^{\ast} - m^{2}\phi^{\ast}\phi - \tilde{\lambda}\vert{\phi}\vert^{4}. \tag{3.3}\label{3.3}
\end{equation}

The complex scalar field of the Lagrangian in \myeqref[Equation~]{3.3} possesses the $SO(2)$ symmetry group as an invariant, which means that it is rotationally invariant in $\mathbb{R}^{2}$. Since the kernel of the real scalar field is the set of all integer rotations, it must be that ker$(\mathit{\phi})$ = $\{2\pi n:n\in \mathbb{Z}\}$. This implies that all solutions must be isomorphic to solutions mapped onto the complex unit circle. Mathematically, this may be expressed as

\begin{equation}
	S^{1} \cong \mathbb{R}^{+} / \mathrm{ker}(\phi) \cong SO(2), \tag{4.1}\label{4.1}
\end{equation}

where

\begin{align}
	S^{1} & = \left\{z \in \mathbb{C}: \vert z \vert = 1\right\}, \tag{4.2.1}\label{4.2.1} \\
	\mathbb{R}^{+} & = \{a \in \mathbb{R} : a > 0\}, \tag{4.2.2}\label{4.2.2} \\
	SO(2) & = \Vast\{\Vast(\begin{matrix} 
		\cos\phi & -\sin\phi \\
		\sin\phi & \cos\phi 
	\end{matrix} \Vast)\hspace*{0.15cm}\Vast \vert \hspace*{0.15cm} \phi \in \mathbb{R} \Vast\}. \tag{4.2.3}\label{4.2.3}
\end{align}

The constraint placed on the gauge parameter is that it must be non-negative. This ensures that the effective potential is a minorant of the field and therefore, creates a stable quantum state that is at its lowest possible energy and is particle-free (a quantum vacuum). Moreover, this ensures that the path integral

\begin{equation}
Z[J] = \int \mathcal{D}\;\phi\exp\Vast\{i\int d^{4}x \Bigg[\frac{1}{2}(\partial_{\mu}\phi \partial^{\mu}\phi
- m^{2}\phi^{2}) - \dfrac{\tilde{\lambda}}{4!}\phi^{4} + J\phi\Bigg]\Vast\}, \tag{5}\label{5}
\end{equation}

over all possible field trajectories, is well defined. See \cite{ref4} for an in depth discussion of quartic interactions for scalar fields and the Feynman path integral formation. For a further discussion of the continuous symmetry groups $SO(2), U(1), SU(2),$ and their $N$-dimensional extensions $SO(N), U(N),$ and $SU(N);$ see \cite{ref5, ref6}.

The $\Phi^{4}$-equation, and more generally, quartic interactions, were first encountered in the studies of phase transitions in statistical mechanics conducted by Landau \cite{Landau1, Landau2, Landau3} in which the energy of bodies changed continuously, but their symmetries changed discontinuously. By the groundbreaking works of Goldstone \cite{Goldstone} and Nambu \cite{Nambu1, Nambu2}, while studying nonperturbative-type `superconductor' solutions in field theories with solutions that were analogous to the Bardeen-Cooper-Schrieffer (BCS) model for superconductors and the proposal of a new `superconductor' model for elementary particles respectively, did the study of quartic interactions enter the realm of particle physics. \\

Obtaining soliton solutions for the $\Phi^{4}$-equation is not novel; several researchers have contributed in this regard. Alquran \textit{et al} \cite{LitRev1} used the Jacobi elliptic expansion method to obtain a previously unknown class of soliton solutions. Similar to the sine-cosine method, delineated in \autoref{SCM} and \autoref{SCM(A)}, the solution to the equation of motion involved expansion, in terms of special function expansions, of sine and / or cosine.

Demiray and Bulut \cite{LitRev2} applied the modified exponential expansion function method to obtain analytical solutions. The method assumed a fractional expansion form for the solution with the numerator and denominator having unequal powers for the exponential function and corresponding coefficients. One then proceeds to algebraically solve for these powers and coefficients.

Akter and Akbar \cite{LitRev3} demonstrated the computational efficiency of the modified simple expansion method by applying it, \textit{inter alia}, to the $\Phi^{4}$ equation. The method involved expanding the solution in terms of the first and second derivatives of some unknown function, which, when solved, yielded solutions in terms of real and complex hyperbolic trigonometric functions.

Triki and Wazwaz \cite{LitRev4} reviewed known kink and anti-kink type soliton solutions for the $\Phi^{4}$-equation and thereafter, studied a generalized form of the equation by raising the field to some arbitrary power. Assuming soliton solutions have hyperbolic secant and tangent forms, bright and dark soliton solutions are obtained via substitution and algebraic manipulation using computer algebra. 

Akbulut \textit{et al} \cite{LitRev9} applied the $\left(\frac{G'}{G}\right)$- and $\left(\frac{1}{G}\right)$-expansion methods to obtain $n$-soliton solutions. Furthermore, they applied Ibragimov's method \cite{Ibragimov1,Ibragimov2,Ibragimov3}, which introduces the differential geometric concepts of the formal Lagrangian and adjoint equations, to obtain conservation laws for the underlying equation.

More general, nonlinear, wave equations have been studied in depth and contribute to the literature on the nonlinear evolution equation, via \textit{ad hoc} techniques, which encapsulate the $\Phi^{4}$-equation and its variants. These include the $\left(\frac{G'}{G}\right)$-expansion method \cite{LitRev5} to the $\left(2+1\right)$-dimensional nonlinear Schr\"{o}dinger equation having dual-power law nonlinearities, the exponential function and modified simple equation methods \cite{LitRev6}  to study the $\left(2+1\right)$-dimensional Zoomeron and Burgers equations, the trial equation method \cite{LitRev7} to study soliton solutions to the generalized Zakharov system of equations, and the generalized Kudryashov method \cite{LitRev8} to obtain soliton solutions to the Davey-Stewartson equation with power-law nonlinearity. \\

In this paper, we calculate soliton solutions to $\Phi^{4}$-equation using the extended hyperbolic tangent and sine-cosine methods. Thereafter, we proceed to obtain conservation laws using the multiplier method in $\left(1+1\right)$-, $\left(2+1\right)$-, and $\left(3+1\right)$-dimensions. Furthermore, we provide a physical interpretation to these conservation laws by discussing their relevance to particle physics. \\

This paper is organized in the following manner:

In \autoref{EOM}, we use the Euler-Lagrange equations to derive the equation of motion by showing how the gauge constant is coupled to the speed of light.

In \autoref{SS}, we apply the extended hyperbolic tangent and sine-cosine methods to obtain soliton solutions. We demonstrate some of the solutions by providing graphical representations of their wave profiles for different parameter values.

In \autoref{CL}, we apply the multiplier method to obtain conservation laws, and summarize them in tabular form for the cases of $\left(1+1\right)$-dimensions.

In \autoref{HDCL}, we apply the multiplier method to obtain conservation laws for the underlying equation in higher dimensions, namely, $\left(2+1\right)$- and $\left(3+1\right)$-dimensions.

In \autoref{DISCUSSION}, a conclusion to this paper is presented wherein the results obtained are reflected upon.

In \autoref{APPENDIX}, a brief discussion of the extended hyperbolic tangent, the sine-cosine, the multiplier, and Lie symmetry methods are presented. This is ascribed to the fact that these techniques are not well known outside the realm of applied mathematics.

\newpage
\section{Equations of Motion}\label{EOM}
For the calculations that follow, we adopt the convention for the Minkowski metric, $\eta_{\mu v}$, to have signature, $(+,-,-,-)$, corresponding to $(+t,-x,-y,-z)$. \\

We use the Euler-Lagrange equation, as derived in \cite{ref2},

\begin{equation}
\partial_{\mu} \Big[\dfrac{\partial \mathscr{L}}{\partial (\partial_{\mu}\phi)}\Big] - \dfrac{\partial \mathscr{L}} {\partial \phi} = 0. \tag{6.1}\label{6.1}
\end{equation}

\enlargethispage{10\baselineskip}
We apply the derivatives to the Lagrangian in \myeqref[Equation~]{1} as follows
\begin{align} 
\ddfrac{\partial\mathscr{L}}{\partial(\partial_{\mu} \phi)} & = \frac{\partial}{\partial\left(\partial_{\mu}\phi\right)}\Bigg(\frac{1}{2}\partial_{\mu} \phi \partial^{\mu}\phi\Bigg) \nonumber \\
& = \frac{1}{2} \ddfrac{\partial}{\partial(\partial_{\mu} \phi)} (\partial_{\mu} \phi \eta^{\mu\nu} \partial_{v}\phi) \nonumber \\
& = \frac{1}{2} \eta^{\mu\nu} \Big[\partial_{v}\phi + \partial_{\mu}\phi \ddfrac{\partial(\partial_{\nu}\phi)}{\partial(\partial_{\mu}\phi)}\Big] \nonumber \\
& = \frac{1}{2} \eta^{\mu\nu} (\partial_{\nu}\phi+\partial_{\mu}\delta^{\nu}_{\mu}) \nonumber \\
& = \frac{1}{2}\eta^{\mu\nu}\partial_{\nu}\phi + \frac{1}{2}\eta^{\mu \nu}\partial_{\mu}\delta_{\mu}^{\nu} \nonumber \\
& = \frac{1}{2}\partial^{\mu}\phi + \frac{1}{2}\partial^{\mu}\phi \nonumber \\
& = \partial^{\mu}\phi. \tag{6.2.1}\label{6.2.1} 
\end{align}

\vspace{-0.9cm}

\begin{align}
\dfrac{\partial\mathscr{L}}{\partial\phi} =&\;\dfrac{\partial}{\partial \phi}\Bigg(-\frac{1}{2}m^{2}\phi^{2} - \frac{\tilde{\lambda}}{4!}\phi^{4}\Bigg) \nonumber \\
=&\; - m^{2}\phi - \frac{\tilde{\lambda}}{3!}\phi^{3}, \tag{6.2.2}\label{6.2.2}
\end{align}
 
where $\delta_{\mu}^{\nu}$ is the Kronecker delta defined as

\begin{equation}  
\delta_{\mu}^{\nu} = 
\begin{cases}
1, \quad \nu = \mu, \\
0, \quad \nu \neq \mu. 
\end{cases} \tag{6.2.3}\label{6.2.3}
\end{equation}

Substituting \myeqref[Equation~]{6.2.1} and \myeqref[Equation~]{6.2.2} into \myeqref[Equation~]{6.1}, we obtain

\begin{equation}
\partial_{\mu}\partial^{\mu}\phi + m^{2}\phi + \frac{\tilde{\lambda}}{3!}\phi^{3} = 0. \tag{6.3}\label{6.3}
\end{equation}

We set the gauge constant, $\tilde{\lambda}$, as being proportional to the speed of light, that is $\tilde{\lambda} \sim c$; this is to ensure that causality is not violated, and the speed of light remains the upper speed limit of the field. We set $\tilde{\lambda} = 6c$ to allow us to cancel the $3!$ in the denominator. Therefore, \myeqref[Equation~]{6.3} becomes

\begin{equation}
\partial_{\mu}\partial^{\mu}\phi + m^{2}\phi + c\phi^{3} = 0. \tag{6.4}\label{6.4}
\end{equation}

In (1+1)-dimensions, \myeqref[Equation~]{6.4} reduces to 

\begin{equation}
\phi_{tt} - \phi_{xx} + m^{2}\phi +c\phi^{3} = 0, \tag{6.5}\label{6.5}
\end{equation}

where $\phi=\phi(x,t)$, and the subscripts denote partial differentiation. \myeqref[Equation~]{6.5} is known as the $\Phi^{4}$-equation, and it becomes the master equation  and the focal point of our study.

\section{Soliton Solutions}\label{SS}

We impose the \textit{ansatz}

\begin{align}
\Xi & = x-wt, \tag{7.1}\label{7.1} \\
\phi(x,t) & = U(\Xi). \tag{7.2}\label{7.2}
\end{align}

\newpage
Thus, \myeqref[Equation~]{6.5} reduces to the ODE

\begin{equation} 
(w^{2}-1) \ddfrac{\mathrm{d}^{2}U}{\mathrm{d}\Xi^{2}} + m^{2}U + cU^{3} = 0. \tag{8}\label{8}
\end{equation}

The hyperbolic tangent and sine-cosine methods are applied to seek solutions to \myeqref[Equation~]{8} in \autoref{SS}.

\subsection{The Extended Hyperbolic Tangent Method}\label{TETM}

Balancing the linear and nonlinear terms, we have $n+2=n=3n$. Thus, with $n=1$, we consider solutions of the form

\begin{equation}
U(\Xi) = a_{0} + a_{1}Z + b_{1}Z^{-1}, \tag{9}\label{9}
\end{equation}

where $Z = \tanh \Xi$. Substituting \myeqref[Equation~]{9} and its associated derivatives into \myeqref[Equation~]{8} and collecting powers of $Z$, we obtain the following system of nonlinear algebraic equations

\begin{align}
Z^{6} & : 2a_{1}w^{2} + a_{1}^{3}c -2a_{1} = 0, \tag{10.1}\label{10.1} \\
Z^{5} & : 3a_{0}a_{1}^{2}c = 0, \tag{10.2}\label{10.2} \\
Z^{4} & : 2a_{1} + 3a_{0}^{2}a_{1}c + 3a_{1}^{2}b_{1}c + a_{1}m^{2} -2a_{1}w^{2} = 0, \tag{10.3}\label{10.3}\\
Z^{3} & : a_{0}^{3}c + 6a_{0}a_{1}b_{1}c + a_{0}m^{2} = 0, \tag{10.4}\label{10.4} \\
Z^{2} & : 2b_{1} + 3a_{0}^{2}b_{1}c + 3a_{1}b_{1}^{2}c + b_{1}m^{2} - 2b_{1}w^{2} = 0, \tag{10.5}\label{10.5} \\
Z^{1} & : 3a_{0}b_{1}^{2}c = 0, \tag{10.6}\label{10.6} \\
Z^{0} & : 2b_{1}w^{2} + b_{1}^{3}c - 2b_{1} = 0. \tag{10.7}\label{10.7}
\end{align}

Solving the system in \myeqref[Equations~]{10.1}-\myeqref[]{10.7} yield the results summarized in \autoref{table:1} of soliton solutions below. 

\FloatBarrier
\begin{table}[p]
	\centering
	\captionsetup{justification=centering}
	\caption{Soliton solutions of the $\Phi^{4}$-equation using the extended hyperbolic tangent method.}
	\label{table:1}
	\vspace*{0.2cm}
		\renewcommand{\arraystretch}{6}
	\begin{tabularx}{0.95\textwidth}{ll}
		\toprule
		\hspace*{1.3cm}\textbf{$\left(a_{0},a_{1},b_{1},w\right)$} & \hspace*{4cm}\textbf{Solution}: $\phi(x,t)$\\
		\hline
		$\Bigg(0,\pm\dfrac{\imath m}{\sqrt{c}},\pm\sqrt{\dfrac{m^{2}+2}{2}}\Bigg)$ & $\phi(x,t) = \pm\dfrac{\imath m}{\sqrt{c}} \tanh \Bigg(x \pm \sqrt{\dfrac{m^{2}+2}{2}} t\Bigg) \ast$ \\
		\hline
		$\Bigg(0, 0, \pm\dfrac{\imath m}{\sqrt{c}},\sqrt{\dfrac{m^{2}+2}{2}}\Bigg)$	& $\phi(x,t) = \pm\dfrac{\imath m}{\sqrt{c}}\coth\Bigg(x \pm \sqrt{\dfrac{m^{2}+2}{2}} t\Bigg) \ast\ast$ \\
		\hline
		$\Bigg(0,-\dfrac{m}{\sqrt{2c}},\dfrac{m}{\sqrt{2c}}, -\ddfrac{\sqrt{4-m^{2}}}{2}\Bigg)$ & $\phi(x,t) = \dfrac{m}{\sqrt{2c}} \tanh\Bigg(x + \dfrac{\sqrt{4-m^{2}}}{2}t\Bigg) + \dfrac{m}{\sqrt{2c}}\coth\Bigg(x + \dfrac{sqrt{4-m^{2}}}{2}t\Bigg)$ \\
		\hline
		$\Bigg(0, \dfrac{m}{\sqrt{2c}}, -\dfrac{m}{\sqrt{2c}}, -\dfrac{\sqrt{4-m^{2}}}{2}\Bigg)$ & $\phi(x,t) = \dfrac{m}{\sqrt{2c}}\tanh\Bigg(x + \dfrac{\sqrt{4-m^{2}}}{2}t\Bigg) - \dfrac{m}{\sqrt{2c}} \coth\Bigg(x + \dfrac{\sqrt{4-m^{2}}}{2}t\Bigg)\ast\ast\ast$\\
		\hline
		$\Bigg(0, -\dfrac{m}{\sqrt{2c}}, \dfrac{m}{\sqrt{2c}}, -\dfrac{\sqrt{4-m^{2}}}{2}\Bigg)$ & $\phi(x,t) = -\dfrac{m}{\sqrt{2c}}\tanh\Bigg(x - \dfrac{\sqrt{4-m^{2}}}{2}t\Bigg) + \dfrac{m}{\sqrt{2c}} \coth\Bigg(x - \dfrac{\sqrt{4-m^{2}}}{2}t\Bigg)$ \\ 
		\hline
		$\Bigg(0, \dfrac{m}{\sqrt{2c}}, -\dfrac{m}{\sqrt{2c}}, -\dfrac{\sqrt{4-m^{2}}}{2}\Bigg)$ & $\phi(x,t) = \dfrac{m}{\sqrt{2c}}\tanh\Bigg(x - \dfrac{\sqrt{4-m^{2}}}{2}t\Bigg) - \dfrac{m}{\sqrt{2c}} \coth\Bigg(x - \dfrac{\sqrt{4-m^{2}}}{2}t\Bigg)$ \\
		\hline
		$\Bigg(0, -\dfrac{\imath m}{2\sqrt{c}}, -\dfrac{\imath m}{2\sqrt{c}}, -\sqrt{\dfrac{m^{2}+8}{8}}\Bigg)$	& $\phi(x,t) = -\ddfrac{\imath m}{2\sqrt{c}} \tanh \Bigg(x + \sqrt{\dfrac{m^{2}+8}{8}}t\Bigg) - \ddfrac{\imath m}{2\sqrt{c}} \coth \Bigg(x + \sqrt{\dfrac{m^{2}+8}{8}}t\Bigg)$ \\
		\hline
		$\Bigg(0, \dfrac{\imath m}{2\sqrt{c}}, \dfrac{\imath m}{2\sqrt{c}}, -\sqrt{\dfrac{m^{2}+8}{8}}\Bigg)$	& $\phi(x,t) = \ddfrac{\imath m}{2\sqrt{c}} \tanh \Bigg(x + \sqrt{\dfrac{m^{2}+8}{8}}t\Bigg) + \ddfrac{\imath m}{2\sqrt{c}} \coth \Bigg(x + \sqrt{\dfrac{m^{2}+8}{8}}t\Bigg)$\\
		\hline
		$\Bigg(0, -\dfrac{\imath m}{2\sqrt{c}}, -\dfrac{\imath m}{2\sqrt{c}}, \sqrt{\dfrac{m^{2}+8}{8}}\Bigg)$	& $\phi(x,t) = -\ddfrac{\imath m}{2\sqrt{c}} \tanh \Bigg(x - \sqrt{\dfrac{m^{2}+8}{8}}t\Bigg) - \ddfrac{\imath m}{2\sqrt{c}} \coth \Bigg(x - \sqrt{\dfrac{m^{2}+8}{8}}t\Bigg)$\\
		\hline
		$\Bigg(0, \dfrac{\imath m}{2\sqrt{c}}, \dfrac{\imath m}{2\sqrt{c}}, \sqrt{\dfrac{m^{2}+8}{8}}\Bigg)$ & $\phi(x,t) = -\ddfrac{\imath m}{2\sqrt{c}} \tanh \Bigg(x - \sqrt{\dfrac{m^{2}+8}{8}}t\Bigg) + \ddfrac{\imath m}{2\sqrt{c}} \coth \Bigg(x - \sqrt{\dfrac{m^{2}+8}{8}}t\Bigg)$ \\
		\bottomrule
	\end{tabularx}
\end{table}
\FloatBarrier


\begin{figure}[H]
	\centering
	\includegraphics[scale=0.6]{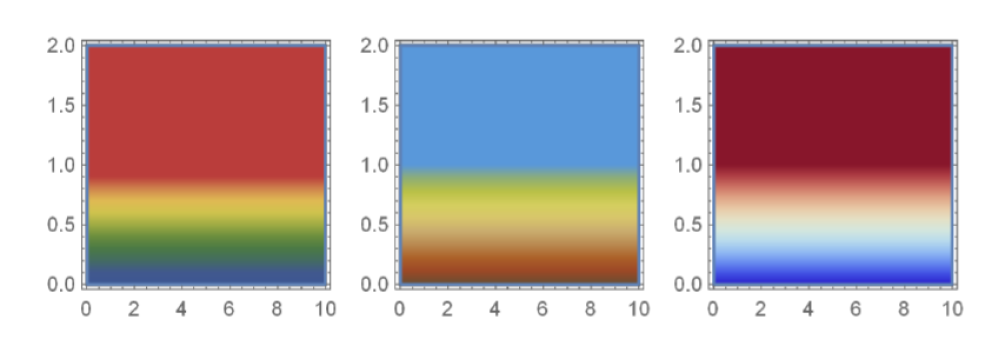}
	\vspace{-0.20cm}
	\caption{Density plot of $\ast$ on the left hand side, $\ast\ast$ in the centre, and $\ast\ast\ast$ on the right hand side with $m=2$ and $c=1$.}
\end{figure}

\subsection{The Sine-Cosine Method}\label{SCM}

We consider solutions of the form

\begin{equation}
U(\Xi) = \lambda \cos^{\beta}(\chi \Xi). \tag{11}\label{11}
\end{equation}

Substituting \myeqref[Equation~]{11} into \myeqref[Equation~]{8} and collecting by powers of $\cos\left({\chi\Xi}\right)$, we get

\begin{align}
& (\beta\lambda\chi^{2} - w^{2}\beta\lambda\chi^{2} - \beta^{2}\lambda\chi^{2} + w^{2}\beta^{2}\lambda\chi^{2})\cos^{\beta - 2}(\chi\Xi) \nonumber \\
& + (m^{2}\lambda + \beta^{2}\lambda\chi^{2} - w^{2}\beta^{2}\lambda\chi^{2})\cos^{\beta}(\chi\Xi) + c\lambda^{3}\cos^{3\beta}(\chi\Xi) = 0. \tag{12}\label{12}
\end{align}

Balancing the powers and equating the coefficients, we obtain the following system of algebraic equations

\begin{align}
& \beta - 2 = 3\beta, \tag{13.1}\label{13.1} \\
& \beta\lambda\chi^{2} - w^{2}\beta\lambda\chi^{2} - \beta^{2}\lambda\chi^{2} + w^{2}\beta^{2}\lambda\chi^{2} =c\lambda^{3}, \tag{13.2}\label{13.2}\\
& m^{2}\lambda + \beta^{2}\lambda\chi^{2} - w^{2}\beta^{2}\lambda\chi^{2} = 0. \tag{13.3}\label{13.3}
\end{align}

Solving the algebraic system in \myeqref[Equations~]{13.1}-\myeqref[]{13.3} simultaneously, while imposing the restriction that $\lambda,\beta \neq 0$, we get

\begin{equation}
(\lambda,\beta,\chi) = \Vast\{\Vast(\pm\sqrt{\frac{2}{c}} m, -1, \pm\ddfrac{m}{\sqrt{w^{2}-1}}\Vast)\Vast\}. \tag{14}\label{14}
\end{equation}

We observe the restriction that $w \neq \pm 1$. Since $\cos(-\theta) = \cos(\theta)$ for an arbitrary value of $\theta$, the solution takes the form

\begin{equation}
U(\Xi) = \pm \sqrt{\frac{2}{c}} m \sec \Bigg(\ddfrac{m}{\sqrt{w^{2}-1}}\Bigg), \tag{15.1}\label{15.1}
\end{equation}

and expressed in terms of the original variables, it is given by

\begin{equation}
\phi(x,t) = \pm \sqrt{\frac{2}{c}} m \sec \Bigg[\dfrac{m}{\sqrt{w^{2}-1}}(x-wt)\Bigg]. \tag{15.2}\label{15.2}
\end{equation}


\begin{figure}[H]
	\centering
	\includegraphics[scale=0.6]{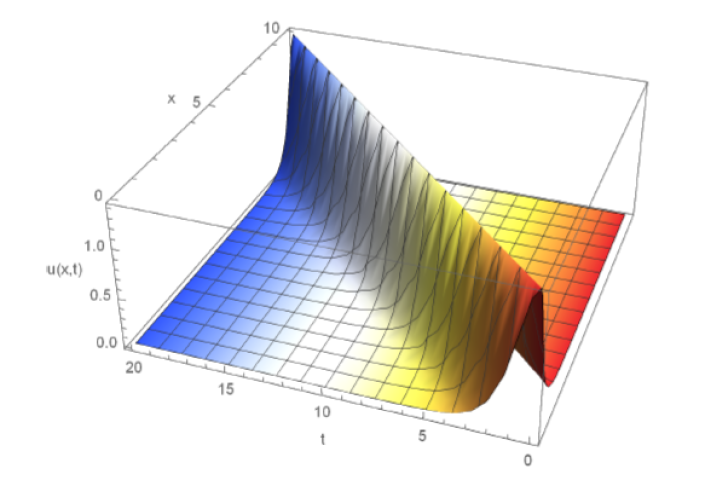}
	\vspace{-0.20cm}
	\caption{Plot of $\phi(x,t)$ with $c=1$, $m=2$ with wave speed, $w=\frac{1}{2}$.}
\end{figure}

\section{Conservation Laws}\label{CL}

We employ the multiplier method to attain the conservation laws. We consider multipliers of the form: $\Lambda = \Lambda(x, t, \phi, \phi_{t}, \phi_{x})$. Thus, the conservation laws take the form

\begin{equation}
\Lambda(x, t, \phi, \phi_{t}, \phi_{x}) [\phi_{tt} - \phi_{xx} + m^{2}\phi +c\phi^{3}] = D_{t}\Sigma^{\left[t\right]} + D_{x}\Sigma^{\left[x\right]}. \tag{16}\label{16}
\end{equation}

\enlargethispage{10\baselineskip}
Applying the Euler-Lagrange derivative in \myeqref[Equation~]{C.1} to \myeqref[Equation~]{16} and separating according to the derivatives of $\phi(x,t)$, we obtain the following set of determining equations

\begin{align}
& \phi_{xt}^{2} : \Lambda_{\phi_{x}\phi_{x}} - \Lambda_{\phi_{t}\phi_{t}} = 0, \tag{17.1}\label{17.1}  \\
& \phi_{xt} : \Lambda_{t\phi_{x}} - \Lambda_{x\phi_{t}} - m^{2}\phi\Lambda_{\phi_{x}\phi_{t}} - c\phi^{3}\Lambda_{\phi_{x}\phi_{t}} = 0, \tag{17.2}\label{17.2} \\
& \phi_{xx} : \Lambda_{t\phi_{t}} - m^{2}\phi\Lambda_{\phi_{x}\phi_{x}} - c\phi^{3}\Lambda_{\phi_{x}\phi_{x}} - 2\Lambda_{\phi} - \Lambda_{x\phi_{x}} = 0, \tag{17.3}\label{17.3} \\
& \phi_{tt} : \Lambda_{t\phi_{t}} - m^{2}\phi\Lambda_{\phi_{t}\phi_{t}} - c\phi^{3}\Lambda_{\phi_{t}\phi_{t}} + 2\Lambda_{\phi} - \Lambda_{x\phi_{x}} = 0, \tag{17.4}\label{17.4} \\
& \phi_{xx}\phi_{tt} : \Lambda_{\phi_{t}\phi_{t}} - \Lambda_{\phi_{x}\phi_{x}} = 0, \tag{17.5}\label{17.5} \\
& \phi_{tt}\phi_{t} : \Lambda_{\phi\phi_{t}} = 0, \tag{17.6}\label{17.6} \\
& \phi_{xx}\phi_{t} : \Lambda_{\phi\phi_{t}} = 0, \tag{17.7}\label{17.7} \\
&\phi_{xt}\phi_{t} : \Lambda_{\phi\phi_{x}} = 0, \tag{17.8}\label{17.8} \\
& \phi_{xt}\phi_{t} : \Lambda_{\phi\phi_{t}} = 0, \tag{17.9}\label{17.9}  \\
& \phi_{tt}\phi_{x} : \Lambda_{\phi\phi_{x}} = 0, \tag{17.10}\label{17.10} \\
& \phi_{xx}\phi_{x} : \Lambda_{\phi\phi_{x}} = 0, \tag{17.11}\label{17.11}  \\
& \phi_{t}^{2} : \Lambda_{\phi\phi} = 0, \tag{17.12}\label{17.12}\\  
& \phi_{x}^{2} : \Lambda_{\phi\phi} = 0, \tag{17.13}\label{17.13}  \\
& \phi_{t} : 2\Lambda_{t\phi} - (m^{2}-3c\phi^{2})\Lambda_{\phi_{t}} - m^{2}\phi\Lambda_{\phi\phi_{t}} - cu^{3}\Lambda_{uu_{t}} = 0, \tag{17.14}\label{17.14} \\
& \phi_{x} = 2\Lambda_{x\phi_{t}} + (m^{2}+3c\phi^{2})\Lambda_{\phi} + (m^{2}\phi +c\phi^{3})\Lambda_{\phi\phi_{x}} = 0, \tag{17.15}\label{17.15} \\
& \phi_{x}^{0} : \Lambda_{tt} + (m^{2}\phi + c\phi^{3})\Lambda_{\phi} - (m^{2}\phi + c\phi^{3})\Lambda_{t\phi_{t}} - (m^{2}\phi + c\phi^{3})\Lambda_{x\phi_{x}} \nonumber \\
&\hspace*{0.7cm} + (m^{2}+3c\phi^{2})\Lambda - \Lambda_{xx} = 0. \tag{17.16}\label{17.16}
\end{align}

\newpage
Solving the system of linear PDEs in \myeqref[Equations~]{17.1}-\myeqref[]{17.16}, we get

\begin{equation}
\Lambda(x, t, \phi, \phi_{t}, \phi_{x}) = (\kappa_{1}x + \kappa_{3})\phi_{t} + (\kappa_{1}t + \kappa_{2})\phi_{x}. \tag{18}\label{18}
\end{equation}

\enlargethispage{10\baselineskip}
where $\kappa_{i} \in \mathbb{R}$ are constants for $1 \leq i \leq 3$. The conservation laws are summarized in   \autoref{table:2} below.

\FloatBarrier
\begin{table}[H]
	\centering
	\caption{Multipliers and conservation laws for the (1+1)-dimensional $\Phi^{4}$-equation.}
	\label{table:2}
	\vspace{0.1cm}
	\renewcommand{\arraystretch}{2.5}
	\begin{tabularx}{0.8\textwidth}{lX}
		\toprule
		\hspace*{0.3cm}\textbf{Multiplier} & \hspace*{2.5cm}\textbf{Conserved Currents} \\
		\hline
		$\Lambda^{\left[1\right]} = x\phi_{t} + t\phi_{x}$	& $\Sigma_{1}^{\left[t\right]} = \frac{1}{4}cx\phi^{4} + \frac{1}{2}m^{2}x\phi^{2} + t\phi_{t}\phi_{x} + \frac{1}{2}x\phi_{t} + \frac{1}{2}x\phi_{x}^{2}$  \\
		& $\Sigma_{1}^{\left[x\right]} = \frac{1}{4}ct\phi^{4} + \frac{1}{2}m^{2}t\phi^{2} - \frac{1}{2}t\phi_{t}^{2} - x\phi_{t}\phi_{x} - \frac{1}{2}t\phi_{x}^{2}$ \\
		\hline
		$\Lambda^{\left[2\right]} = \phi_{x}$ & $\Sigma_{2}^{\left[t\right]}$ = $\phi_{t}\phi_{x}$  \\
		& $\Sigma_{2}^{\left[x\right]} = \frac{1}{4}c\phi^{4} + \frac{1}{2}m^{2}\phi^{2} - \frac{1}{2}\phi_{t}^{2} - \frac{1}{2}\phi_{x}^{2}$ \\
		\hline
		$\Lambda^{\left[3\right]} = \phi_{t}$ & $\Sigma^{\left[t\right]}_{3}=\frac{1}{2}\phi_{x}^{2} + \frac{1}{4}c\phi^{4} + \frac{1}{2}m^{2}\phi^{2} + \frac{1}{2}\phi_{t}^{2}$ \\
		& $\Sigma_{3}^{\left[x\right]} = -\phi_{t}\phi_{x}$ \\
		\bottomrule
	\end{tabularx}
\end{table}
\FloatBarrier

\section{Higher-dimensional Conservation Laws}\label{HDCL}

In (2+1)-dimensions, \myeqref[Equation~]{6.4} becomes

\begin{equation}
\phi_{tt} - \phi_{xx} - \phi_{yy} + m^{2}\phi + c\phi^{3} = 0, \tag{19}\label{19}
\end{equation}

where $\phi=\phi(x,y,t)$. We consider multipliers of the form: $\Lambda = \Lambda(x,y,t,\phi,\phi_{t}, \phi_{x}, \phi_{y})$. Thus, the conservation laws take the form

\vspace{-0.5cm}

\begin{equation}
\Lambda = \Lambda(x,y,t,\phi,\phi_{t}, \phi_{x}, \phi_{y})[\phi_{tt} - \phi_{xx} - \phi_{yy} + m^{2}\phi + c\phi^{3}] = D_{t}\Sigma^{\left[t\right]} + D_{x}\Sigma^{\left[x\right]} + D_{y}\Sigma^{\left[y\right]}. \tag{20}\label{20}
\end{equation}

Henceforth, the calculations become monotonous, and therefore, we use $\texttt{MATHEMATICA}$ \cite{ref7} to circumvent these long and convoluted calculations. In doing so, we find that the general form of the multiplier is

\begin{align}
\Lambda(x,y,t,\phi,\phi_{t}, \phi_{x}, \phi_{y}) =&\; (\kappa_{5}y + \kappa_{6}x + \kappa_{9})u_{t} + (\kappa_{4}y + \kappa_{6}t + \kappa_{7})u_{x} \nonumber \\
&\; - (\kappa_{4}x - \kappa_{5}t - \kappa_{8})u_{y}, \tag{21}\label{21}
\end{align}

where $\kappa_{i} \in \mathbb{R}$ are constants for $4 \leq i \leq 9$. The conservation laws are summarized in \autoref{table:3} below.

\FloatBarrier
\begin{table}[p]
	\centering
	\caption{Multipliers and conservation laws for the (2+1)-dimensional $\Phi^{4}$-equation.}
	\label{table:3}
	\vspace{0.1cm}
	\renewcommand{\arraystretch}{3.52}
	\begin{tabularx}{\textwidth}{lX}
		\toprule
		\hspace*{0.3cm}\textbf{Multiplier} & \hspace*{4.5cm}\textbf{Conserved Currents} \\
		\hline
		$\Lambda^{\left[4\right]} = y\phi_{x} - x\phi_{y}$	& $\Sigma^{\left[t\right]}_{4} = y\phi_{t}\phi_{x} - x\phi_{t}\phi_{y}$ \\ 
		&  $\Sigma_{4}^{\left[x\right]} = \frac{1}{2}m^{2}y\phi^{2} + \frac{1}{4}cy\phi^{4} - \frac{1}{2}y\phi_{x}^{2} + \frac{1}{2}y\phi_{y}^{2} - \frac{1}{2}y\phi_{t}^{2} - \frac{1}{2}cx^{2}\phi^{3}\phi_{y} \newline 
		\hspace*{0.5cm}	-\frac{1}{2}m^{2}x^{2}\phi\phi_{y} + x\phi_{x}\phi_{y}$ \\
		& $\Sigma_{4}^{\left[y\right]} = -y\phi_{x}\phi_{y} + \frac{1}{2}cx^{2}\phi^3\phi_{x} + \frac{1}{2}m^{2}x^{2}\phi\phi_{x} + \frac{1}{2}x\phi_{y}^{2} - \frac{1}{2}x\phi_{x}^{2} + \frac{1}{2}x\phi_{t}^{2}$ \\
		\hline
		$\Lambda^{\left[5\right]} = t\phi_{y} - y\phi_{t}$	& $\Sigma_{5}^{\left[t\right]} = t\phi_{t}\phi_{y} + \frac{1}{2}y\phi_{y}^{2} + \frac{1}{2}y\phi_{t}^{2} + \frac{1}{4}cy\phi^{4} + \frac{1}{2}m^{2}y\phi^{2} + \frac{1}{2}y\phi_{x}^{2}$ \\ 
		& $\Sigma_{5}^{\left[x\right]} = ctx\phi^{3}\phi_{y} + m^{2}tx\phi\phi_{y} - t\phi_{x}\phi_{y} - y\phi_{t}\phi_{x}$ \\
		& $\Sigma_{5}^{\left[y\right]} = -y\phi_{t}\phi_{y} - ctx\phi^{3}\phi_{x} - m^{2}tx\phi\phi_{x} - m^{2}tx\phi\phi_{x} - \frac{1}{2}t\phi_{y}^{2} + \frac{1}{2}t\phi_{x}^{2}$ \\
		\hline
		$\Lambda^{\left[6\right]} = t\phi_{x} - x\phi_{t}$	& $\Sigma_{6}^{\left[t\right]} = \frac{1}{2}x\phi_{y}^{2} + \frac{1}{2}x\phi_{t}^{2} + t\phi_{t}\phi_{x} + \frac{1}{4}cx\phi^{4} + \frac{1}{2}m^{2}x\phi^{2} + \frac{1}{2}x\phi_{x}^{2}$ \\
		& $\Sigma_{6}^{\left[x\right]} = \frac{1}{2}m^{2}t\phi^{2} + \frac{1}{4}ct\phi^{4} - \frac{1}{2}t\phi_{x}^{2} + \frac{1}{2}t\phi_{y}^{2} - \frac{1}{2}t\phi_{t}^{2} - x\phi_{t}\phi_{x}$ \\
		& $\Sigma_{6}^{\left[y\right]}$ = $-t\phi_{x}\phi_{y} - x\phi_{t}\phi_{y}$ \\
		\hline
		$\Lambda^{\left[7\right]} = \phi_{x}$ & $\Sigma_{7}^{\left[t\right]} = \phi_{t}\phi_{x} $ \\
		& $\Sigma_{7}^{\left[x\right]} = \frac{1}{4}c\phi^{4} + \frac{1}{2}m^{2}\phi^{2} - \frac{1}{2}\phi_{x}^{2} + \frac{1}{2}\phi_{y}^{2} - \frac{1}{2}\phi_{t}^{2}$ \\
		& $\Sigma_{7}^{\left[y\right]} = -\phi_{x}\phi_{y}$\\
		\hline
		$\Lambda^{\left[8\right]} = \phi_{y}$ & $\Sigma_{8}^{\left[t\right]} = \phi_{y}\phi_{t} $  \\
		&  $ \Sigma_{8}^{\left[x\right]} = cx\phi^{3}\phi_{y} + m^{2}x\phi\phi_{y} -\phi_{x}\phi_{y}$ \\
		& $ \Sigma_{8}^{\left[y\right]} = -\frac{1}{2}\phi_{t}^{2} - c\phi^{3}x\phi_{x} - m^{2}x\phi\phi_{x} + \frac{1}{2}\phi_{x}^{2} - \frac{1}{2}\phi_{y}^{2}$ \\
		\hline
		$\Lambda^{\left[9\right]} = \phi_{t}$ & $\Sigma_{9}^{\left[t\right]} = \frac{1}{4}c\phi^{4} + \frac{1}{2}m^{2}\phi^{2} + \frac{1}{2}\phi_{t}^{2} + \frac{1}{2}\phi_{x}^{2} + \frac{1}{2}\phi_{y}^{2}$  \\
		& $\Sigma_{9}^{\left[x\right]} = -\phi_{t}\phi_{x}$ \\
		& $\Sigma_{9}^{\left[y\right]} = -\phi_{t}\phi_{y}$ \\
		\bottomrule
	\end{tabularx}
\end{table}
\FloatBarrier

In (3+1)-dimensions, \myeqref[Equation~]{6.4} becomes 

\begin{equation}
\phi_{tt} - \phi_{xx} - \phi_{yy} - \phi_{zz} + m^{2}\phi + c\phi^{3} = 0, \tag{22}\label{22}
\end{equation}

where $\phi=\phi(x,y,z,t)$. We consider multipliers of the form: $\Lambda = \Lambda(x,y,z,t,\phi_{t},\phi_{x},\phi_{y},\phi_{z})$. Thus, the conservation laws take the form

\begin{align}
& \Lambda(x,y,z,t,\phi_{t},\phi_{x},\phi_{y},\phi_{z})[\phi_{tt} - \phi_{xx} - \phi_{yy} - \phi_{zz} + m^{2}\phi + c\phi^{3}] \nonumber \\
& = D_{t}\Sigma^{\left[t\right]} + D_{x}\Sigma^{\left[x\right]} + D_{y}\Sigma^{\left[y\right]} + D_{z}\Sigma^{\left[z\right]}. \tag{23}\label{23}
\end{align}

Employing $\texttt{MATHEMATICA}$ \cite{ref7}, we find that the general form of the multiplier is

\begin{align}
\Lambda(x,y,z,t,\phi_{t},\phi_{x},\phi_{y},\phi_{z}) =&\; (\kappa_{12}y + \kappa_{14}x + \kappa_{15}z + \kappa_{19})u_{t} + (\kappa_{10}y + \kappa_{13}z + \kappa_{14}t + \kappa_{16} )\phi_{x} \nonumber \\
& \;- (\kappa_{10}x - \kappa_{11}z - \kappa_{12}t - \kappa_{17} )\phi_{y} \nonumber  \\
& \;+ (\kappa_{11}y + \kappa_{13}x + \kappa_{15}t + \kappa_{18} )\phi_{z}, \tag{24}\label{24}
\end{align}

where $\kappa_{i} \in \mathbb{R}$ are constants for $10 \leq i \leq 19$. The conservation laws are summarized in \autoref{table:4} - \autoref{table:6} below.

\FloatBarrier
\begin{table}[p!]
	\centering
	\caption{Multipliers and conservation laws for the (3+1)-dimensional $\Phi^{4}$-equation -- Part I.}
	\label{table:4}
	\vspace{0.1cm}
	\renewcommand{\arraystretch}{3.9}
	\begin{tabularx}{\textwidth}{lX}
		\toprule
		\hspace*{0.5cm}\textbf{Multiplier} & \hspace{4.5cm}\textbf{Conserved Currents} \\
		\hline
		$\Lambda^{\left[10\right]} = y\phi_{x} - x\phi_{y}$ & $\Sigma_{10}^{\left[t\right]} = y\phi_{t}\phi_{x} - x\phi_{t}\phi_{y}$ \\ 
		& $\Sigma_{10}^{\left[x\right]} = x\phi_{x}\phi_{y} - \frac{1}{2}y\phi_{t}^{2} + \frac{1}{2}y\phi_{z}^{2} + \frac{1}{2}y\phi_{y}^{2} - \frac{1}{2}cx^{2}\phi^{3}\phi_{y} - \frac{1}{2}m^{2}x^{2}\phi\phi_{y} - 
		\frac{1}{2}y\phi_{x} ^{2} \newline  \hspace*{0.5cm} + \frac{1}{2}m^{2}y\phi^{2} + \frac{1}{4}cy\phi^{4}$ \\
		& $\Sigma_{10}^{\left[y\right]} = \frac{1}{2}x\phi_{y}^{2} - \frac{1}{2}x\phi_{z}^{2} + \frac{1}{2}x\phi_{t}^{2} - \frac{1}{2}x\phi_{x}^{2} + \frac{1}{2}cx^{2}\phi^{3}u_{x} + \frac{1}{2}m^{2}x^{2}\phi\phi_{x} - y\phi_{x}\phi_{y}$ \\ 
		& $\Sigma_{10}^{\left[z\right]} = -y\phi_{x}\phi_{z} + x\phi_{y}\phi_{z}$\\
		\hline
		$\Lambda^{\left[11\right]} = -z\phi_{y} - y\phi_{z}$ & $\Sigma_{11}^{\left[t\right]} = -z\phi_{t}\phi_{y} + y\phi_{t}\phi_{z}$  \\
		&  $\Sigma_{11}^{\left[x\right]} = -cxz\phi^{3}\phi_{y} - m^{2}xz\phi\phi_{y} + z\phi_{x}\phi_{y} - y\phi_{x}\phi_{z}$\\
		& $\Sigma_{11}^{\left[y\right]} = \frac{1}{2}z\phi_{y}^{2} - \frac{1}{2}z\phi_{z}^{2} + \frac{1}{2}z\phi_{t} - \frac{1}{2}z\phi_{x}^{2} + cxz\phi^{3}\phi_{x} + m^{2}xz\phi\phi_{x} + \frac{1}{2}cy^{2}\phi^{3}\phi_{z} \newline \hspace*{0.5cm} + \frac{1}{2}m^{2}\phi^{2}\phi\phi_{z} - y\phi_{y}\phi_{z}$  \\
		& $\Sigma_{11}^{\left[z\right]} = \frac{1}{2}y\phi_{x}^{2} + \frac{1}{2}y\phi_{y}^{2} - \frac{1}{2}y\phi_{z}^{2} - \frac{1}{2}y\phi_{t}^{2} - \frac{1}{2}cy^{2}\phi^{3}\phi_{y} - \frac{1}{2}m^{2}y^{2}\phi\phi_{y} + z\phi_{y}\phi_{z}$ \\
		\hline
		$\Lambda^{\left[12\right]} = t\phi_{y} + y\phi_{t}$ & $\Sigma_{12}^{\left[t\right]} = \frac{1}{2}cy\phi^{4} + \frac{1}{2}m^{2}y\phi^{2} + \frac{1}{2}y\phi_{t}^{2} + \frac{1}{2}y\phi_{y}^{2} + \frac{1}{2}y\phi_{x} + t\phi_{t}\phi_{y} + \frac{1}{2}y\phi_{z}^{2}$ \\
		& $\Sigma_{12}^{\left[x\right]} = ctx\phi^{3}\phi_{y} + m^{2}tx\phi\phi_{y} - t\phi_{x}\phi_{y} - y\phi_{t}\phi_{x}$ \\
		& $\Sigma_{12}^{\left[y\right]} = -\frac{1}{2}t\phi_{y}^{2} + \frac{1}{2}t\phi_{z}^{2} - \frac{1}{2}t\phi_{t}^{2} + \frac{1}{2}t\phi_{x} - ctx\phi^{3}\phi_{x} - m^{2}tx\phi\phi_{x} - y\phi_{t}\phi_{y}$ \\
		& $\Sigma_{12}^{\left[z\right]} = -t\phi_{y}\phi_{z} - y\phi_{t}\phi_{z}$ \\
		\hline
		$\Lambda^{\left[13\right]} = -z\phi_{x} + x\phi_{z}$ & $\Sigma_{13}^{\left[t\right]}	= -z\phi_{t}\phi_{x} + x\phi_{t}\phi_{z}$ \\
		& $\Sigma_{13}^{\left[x\right]} = -x\phi_{x}\phi_{z} + \frac{1}{2}z\phi_{t}^{2} - \frac{1}{2}z\phi_{z}^{2} - \frac{1}{2}z\phi_{y}^{2} + \frac{1}{2}z\phi_{x}^{2} - \frac{1}{4}cz\phi^{4} - \frac{1}{2}m^{2}z\phi^{2}$ \\
		& $\Sigma_{13}^{\left[y\right]}$ =  $cxy\phi^{3}u_{z} + m^{2}xy\phi\phi_{z} + z\phi_{x}\phi_{y} - x\phi_{y}\phi_{z}$ \\
		& $\Sigma_{13}^{\left[z\right]} $ = $-\frac{1}{2}x\phi_{t}^{2} + \frac{1}{2}x\phi_{x}^{2} + \frac{1}{2}x\phi_{y}^{2} - \frac{1}{2}x\phi_{z}^{2} + z\phi_{x}\phi_{z} - cxy\phi^{3}\phi_{y} - m^{2}xy\phi\phi_{y}$ \\
		\bottomrule
	\end{tabularx}
\end{table}
\FloatBarrier

\FloatBarrier
\begin{table}[p!]
	\centering
	\caption{Multipliers and conservation laws for the (3+1)-dimensional $\Phi^{4}$-equation -- Part II.}
	\label{table:5}
	\vspace{0.2cm}
	\renewcommand{\arraystretch}{3.9}
	\begin{tabularx}{\textwidth}{lX}
		\toprule
		\hspace*{0.3cm}\textbf{Multiplier} & \hspace*{5cm}\textbf{Conserved Currents} \\
		\hline
		$\Lambda^{\left[14\right]} = t\phi_{x} + x\phi_{t}$ & $\Sigma_{14}^{\left[t\right]} = \frac{1}{4}cx\phi^{4} + \frac{1}{2}m^{2}x\phi^{2} + \frac{1}{2}x\phi_{t}^{2} + \frac{1}{2}x\phi_{y}^{2} + \frac{1}{2}x\phi_{x}^{2} + t\phi_{t}\phi_{x} + \frac{1}{2}x\phi_{z}^{2}$ \\ 
		& $\Sigma_{14}^{\left[x\right]} = -x\phi_{t}\phi_{x} - \frac{1}{2}t\phi_{t}^{2} + \frac{1}{2}t\phi_{z}^{2} + \frac{1}{2}t\phi_{y}^{2} - \frac{1}{2}t\phi_{x}^{2} + \frac{1}{4}ct\phi^{4} + \frac{1}{2}m^{2}t\phi^{2}$ \\
		& $\Sigma_{14}^{\left[y\right]} = -t\phi_{x}\phi_{y} - x\phi_{t}\phi_{y}$ \\ 
		& $\Sigma_{14}^{\left[z\right]} = -t\phi_{x}\phi_{z} - x\phi_{t}\phi_{z}$ \\
		\hline
		$\Lambda^{\left[15\right]} = t\phi_{z} + z\phi_{t}$ & $\Sigma_{15}^{\left[t\right]} = \frac{1}{2}m^{2}z\phi^{2} + \frac{1}{4}cz\phi^{4} + \frac{1}{2}z\phi_{t}^{2}+ \frac{1}{2}z\phi_{y}^{2} + \frac{1}{2}z\phi_{x}^{2} + \frac{1}{2}z\phi_{z}^{2} + t\phi_{t}\phi_{z}$ \\
		& $\Sigma_{15}^{\left[x\right]} = -t\phi_{x}\phi_{z} - z\phi_{t}\phi_{x}$ \\
		& $\Sigma_{15}^{\left[y\right]} = cty\phi^{3}\phi_{z} + m^{2}ty\phi\phi_{z} - t\phi_{y}\phi_{z} - z\phi_{t}\phi_{y}$ \\
		& $\Sigma_{15}^{\left[z\right]} = -\frac{1}{2}t\phi_{t}^{2} + \frac{1}{2}t\phi_{x}^{2} + \frac{1}{2}t\phi_{y}^{2} - \frac{1}{2}t\phi_{z}^{2} - z\phi_{t}\phi_{z} - cty\phi^{3}\phi_{y} - m^{2}ty\phi\phi_{y}$ \\
		\hline
		$\Lambda^{\left[16\right]} = \phi_{x}$	& $\Sigma_{16}^{\left[t\right]} = \phi_{t}\phi_{x}$ \\
		& $\Sigma_{16}^{\left[x\right]} = \frac{1}{4}c\phi^{4} + \frac{1}{2}m^{2}\phi^{2} + \frac{1}{2}\phi_{z}^{2} - \frac{1}{2}\phi_{x}^{2} -\frac{1}{2}\phi_{t}^{2} + \frac{1}{2}\phi_{y}^{2}$  \\
		& $\Sigma_{16}^{\left[y\right]} = -\phi_{x}\phi_{y}$ \\
		& $\Sigma_{16}^{\left[z\right]} = -\phi_{x}\phi_{z}$ \\
		\hline
		$\Lambda^{\left[17\right]} = \phi_{y}$	&  $\Sigma_{17}^{\left[t\right]} = \phi_{y}\phi_{t}$ \\
		& $\Sigma_{17}^{\left[x\right]} = cx\phi^{3}\phi_{y} + m^{2}x\phi\phi_{y} - \phi_{x}\phi_{y}$ \\
		& $\Sigma_{17}^{\left[y\right]} = -\frac{1}{2}\phi_{y}^{2} + \frac{1}{2}\phi_{z}^{2} - \frac{1}{2}\phi_{t}^{2} + \frac{1}{2}\phi_{x} - cx\phi^{3}u_{x} - m^{2}x\phi\phi_{x}$ \\
		& $\Sigma_{17}^{\left[z\right]} = -\phi_{y}\phi_{z}$ \\
		\bottomrule
	\end{tabularx}
\end{table}
\FloatBarrier

\FloatBarrier
 \begin{table}[p]
	\centering
	\caption{Multipliers and conservation laws for the (3+1)-dimensional $\Phi^{4}$-equation -- Part III.}
	\label{table:6}
	\vspace{0.2cm}
	\renewcommand{\arraystretch}{3}
	\begin{tabularx}{\textwidth}{lX}
		\toprule
		\textbf{Multiplier} & \hspace*{5cm} \textbf{Conserved Currents} \\
		\hline
		$\Lambda^{\left[18\right]} = u_{z}$	& $\Sigma_{18}^{\left[t\right]} = \phi_{t}\phi_{z}$ \\ 
		& $\Sigma_{18}^{\left[x\right]} = -\phi_{x}\phi_{z}$  \\
		& $\Sigma_{18}^{\left[y\right]} = cy\phi^{3}u_{z} + m^{2}y\phi\phi_{z} - \phi_{y}\phi_{z}$ \\ 
		& $\Sigma_{18}^{\left[z\right]} = \frac{1}{2}\phi_{y}^{2} - \frac{1}{2}\phi_{z}^{2} - \frac{1}{2}\phi_{t}^{2} + \frac{1}{2}\phi_{x}^{2} - cy\phi^{3}\phi_{y} - m^{2}y\phi\phi_{y}$ \\
		\hline
		$\Lambda^{\left[19\right]} = \phi_{t}$	&  $\Sigma_{19}^{\left[t\right]} = \frac{1}{4}c\phi^{4} + \frac{1}{2}m^{2}\phi^{2} + \frac{1}{2}\phi_{t}^{2} + \frac{1}{2}\phi_{y}^{2} + \frac{1}{2}u_{x}^{2}  + \frac{1}{2}\phi_{z}^{2}$ \\
		& $\Sigma_{19}^{\left[x\right]} = -\phi_{t}\phi_{x}$ \\
		& $\Sigma_{19}^{\left[y\right]} = -\phi_{t}\phi_{x}$ \\
		& $\Sigma_{19}^{\left[z\right]} = -\phi_{t}\phi_{z}$ \\
		\bottomrule
	\end{tabularx}
\end{table}
\FloatBarrier

\section{Conclusion}\label{DISCUSSION}

In this paper, we provided an in depth study of the interacting scalar fields with the Lagrangian extended to quartic powers of the field. Using the Euler-Lagrange equations, the master equation of motion was derived for a real scalar field. By coupling the gauge constant with the speed of light and astutely selecting the proportionality constant, it was shown how the cubic field term could be stated in terms of the speed of light. The properties of the field were reviewed, particularly the parity invariance. Thereafter, a synopsis of a two-field extension of the Lagrangian was provided, and thereafter, by adopting a convenient mathematical \textit{ansatz}, it was shown how the two-field Lagrangian can be stated in terms of a scalar field. The two-dimensional rotational invariance for the complex scalar field Lagrangian was then shown, and using the field's kernel, it was shown how solutions can be mapped to the unit circle in the complex plane. The exigent reason for this was to show that quartic interactions can be considered for both real and complex scalar fields. An insightful discussion about the nature of the gauge term, accompanied by further references to ancillary material, was provided. \\

The equations of motion were then solved in terms of the soliton waves using the extended hyperbolic tangent method and the sine-cosine method. In the former case, ten new solutions were produced. The nature of the solutions were simply a linear combination of the complex hyperbolic tangent and cotangent functions. Physical representations of the solutions were produced, for three cases, by taking the magnitude of the fields and constructing density plots. The darker hues in the density plot show a larger value of the field, and conversely, a lighter hue shows a smaller value of the field. In the latter case, two new solutions were produced, varying only by sign. If one follows through on the mathematics, it is evident that four solutions ought to be produced. However, since the secant function is positive for all values of the argument, the additional solutions merge. An interesting feature of these solutions is that they lie on the real line. A plot of the positive solution, for a specific choice of the constants, yields a cuspon solitory wave for the domain, $(x,t) \in [0,10] \times [0,20]$. The divergence of the wave's derivative is clearly evident; the wave monotonically increases from zero to its maximum value, $u_{\mathrm{max}} = 1$, and then decreases to zero. It thus begs the question: \textit{"Symmetry methods play a central role in this study, so why are the equations of motion not studied in terms of this approach?"}. If one undertakes the arduous task of determining Lie point symmetries, one would find that for the (1+1)-dimensional case, a three-dimensional sub-algebra is spanned by the vector fields

\begin{align}
\Gamma_{1} & = \partial_{x}, \tag{25.1}\label{25.1} \\ 
\Gamma_{2} & = \partial_{t}, \tag{25.2}\label{25.2} \\
\Gamma_{3} & = t\partial_{x} + x\partial_{t}. \tag{25.3}\label{25.3}
\end{align}

with the algebraic structure summarized in \autoref{table:7} below.

\FloatBarrier
\begin{table}[H]
	\centering
	\caption{Soliton solutions of the $\Phi^{4}$-equation using the extended hyperbolic tangent method.}
	\label{table:7}
	\vspace*{0.2cm}
	\renewcommand{\arraystretch}{2.3}
	\begin{tabularx}{0.30\textwidth}{llll}
		\toprule
		\hline
		$\boldsymbol{\llbracket \Gamma_{i},\Gamma_{j}\rrbracket}$ & $\boldsymbol{\Gamma_{1}}$ & $\boldsymbol{\Gamma_{2}}$ & $\boldsymbol{\Gamma_{3}}$ \\ 
		\hline
		$\boldsymbol{\Gamma_{1}}$ & 0 & 0 & $\Gamma_{2}$ \\ 
		\hline
		$\boldsymbol{\Gamma_{2}}$ & 0 & 0 & $\Gamma_{1}$ \\ 
		\hline
		$\boldsymbol{\Gamma_{3}}$ & $-\Gamma_{2}$ & $-\Gamma_{1}$ & 0 \\
		\hline
		\bottomrule 
	\end{tabularx}
\end{table}
\FloatBarrier

Descriptively, and due to the nature of Lie point symmetries, the equation is spatially, temporally, and Galilean invariant. The Galilean invariance is not coincidental; a requirement of constructing a reasonable quantum field theory is that the equations of motion must be invariant under Galilean boosts; see \cite{ref2} for a comprehensive discussion. From \autoref{table:7}, we observe that if $\Gamma_{1}$ is used, a solution in the spatial plane will be obtained, which is uninteresting. Furthermore, a linear combination of $\Gamma_{1}$ and $\Gamma_{2}$ naturally imply travelling wave solutions; hence, this justifies initially studying the solution to \myeqref[Equation~]{6.5} in terms of solitary wave theory. Additionally, we detect that $\Gamma_{3}$ cannot be used for constructing solutions. \\

Lastly, using the multiplier method, conservation laws for the equations of motion are constructed. We adopt this approach because Ibragimov's method \cite{Ibragimov1,Ibragimov2,Ibragimov3} fails due the lackluster nature of the Lie point symmetries, and we hypothesize that even the adjoint equations would possess prosaic Lie point symmetries. Initially, conservation laws are constructed in the (1+1)-dimensional case, whereby, three conservation laws result from three disparate multipliers and thereafter, extend to the (2+1)-dimensional case, whereby, five conservation laws result from five distinctive multipliers, and the (3+1)-dimensional case, whereby, ten conservation laws result from ten particular multipliers. In all three cases we observed that the integral of motion is non-trivial. We note that the multipliers extend up to first-order derivatives of the field, always involving the temporal variation, and corresponding spatial variation, interrelated to dimensionality. An interpretation of the conservation laws, with respect to particle physics, is that of quantum numbers. Since the conservation laws describe the dynamics of the evolving field and what quantities remain constant in the motion thereof, we can say that they present eigenvalues of operators commensurate with the Hamiltonian. Since, in the (1+1)-, (2+1)-, and (3+1)-dimensional cases, we know all trivial and nontrivial conservation laws, we may therefore characterize this as the basis state of the interacting scalar fields and consequently, determine the conserved quantities together. Moreover, these conservation laws can be used to categorize the localized internal (flavour) symmetries (electric charge, lepton number, baryon number, isopin, charm, and strangeness) and spacetime symmetries (Poincar\'e symmetry, related to time reversal; spin, related to rotational symmetry; charge parity, related to charge conjugation and parity transformations - sign flips in the spatial dimensions) of the interacting fields. 

\section{Appendix}\label{APPENDIX}

\subsection{The Extended Hyperbolic Tangent (XTanh) Method}\label{TEHTM}

The Extended Hyperbolic Tangent method \cite{Xtanh1, Xtanh2, Xtanh3} is an extension of the hyperbolic tangent (Tanh) method proposed by Malfliet and his collaborator, Hereman \cite{Tanh1, Tanh2, Tanh3, Tanh4}. Wadati \cite{Wadati1, Wadati2, Wadati3} defines a \textit{soliton} as a nonlinear wave that has the following properties: 

\begin{enumerate}
	\item A localized wave that propagates without changing of its physical properties (shape,
	velocity, frequency, amplitude, wavelength, etc.). 
	\item It has the particle property; it is a localized wave that has infinite support or
	exponential tails.
\end{enumerate}

More technically, solitons are waves that are stable against mutual collisions, and they retain their identity.

We outline the extended hyperbolic tangent method, algorithmically, for finding soliton solutions, below.

\begin{enumerate}
	\item Consider the system of PDEs of the form
	\begin{equation}
	H(x,t,\phi_{x},\phi_{t},\phi_{xx},\phi_{tt},\ldots)=0. \tag{A.1}\label{A.1}
	\end{equation}
	Introduce the variable
	\begin{equation} 
	\Xi=\mathbf{x}-wt, \tag{A.2.1}\label{A.2.1}
	\end{equation}
	where $w<\infty$ is the velocity of the wave profile and
	\begin{equation}
	Z=\tanh\Xi. \tag{A.2.2}\label{A.2.2}
	\end{equation}
	\item Convert the system of nonlinear PDEs in \myeqref[Equation~]{A.1} into the nonlinear system of ODEs
	\begin{equation}
	K_{m}\left(V,\frac{dV}{d\Xi},\frac{d^{2}V}{d\Xi^{2}},\ldots\right)=0. \tag{A.3}\label{A.3}
	\end{equation}
	\item Consider solutions of the finite power series form
	\begin{equation}
	V(\Xi)=\sum_{i=0}^{n}a_{i}Z^{i}+\sum\limits_{i=1}^{n}b_{i}Z^{-i}, \tag{A.4}\label{A.4}
	\end{equation}
	where $a_{i}, b_{i}\in\mathbb{R}$ are constants to be determined.
	\item Determine the degree of the polynomial solution in $Z$. Equate every pair of possible highest exponents in the equation to get a linear system for $n$. Reject all solutions where $n\notin\mathbb{Z}^{+}$. We delineate the rules for assigning a polynomial degree, according to powers of the derivatives of the dependent variable, in \autoref{table:8} below.
	\begin{table}[H]
		\centering
		\caption{Rules for assigning polynomial degrees}
		\label{table:8}
		\renewcommand{\arraystretch}{2.3}
		\begin{tabular}{ll}
			\toprule
			\hline
			\textbf{Functional Form} &\textbf{Polynomial Degree} \\
			\hline
			$V(\Xi)$ &$n$ \\
			\hline
			$\frac{dV}{d\Xi}$ &$n+1$ \\
			\hline
			$\frac{d^{2}V}{d\Xi^{2}}$ &$n+2$ \\
			\hline
			\hspace*{0.2cm}\vdots &\hspace*{0.4cm}\vdots \\
			\hline
			$\frac{d^{r}V}{d\Xi^{r}}$ &$n+r$ \\
			\hline
			$\left(\frac{d^{r}V}{d\Xi^{r}}\right)^{q}$ &$q\left(n+r\right)$ \\
			\hline
			\bottomrule
		\end{tabular}	
	\end{table} 
	\item Derive, and solve the system of nonlinear algebraic equations for the coefficients $a_{i}$ and $b_{i}$. \\
	\item Back substitute the solutions for $a_{i}$ and $b_{i}$ into the expansion in \myeqref[Equation~]{A.4}, and obtain the explicit solution in terms of the original variables $\left(x,t,\phi\right)$. \\
	\item Test the final solution from Step 6 to ascertain its precision.
\end{enumerate}

\subsection{The Sine-Cosine Method}\label{SCM(A)}

The Sine-cosine method was introduced by Wazwaz \cite{Wazwaz1, Wazwaz2} for finding travelling wave solutions of generalized KdV-type equations. We outline, algorithmically, the sine-cosine method for finding soliton solutions, below.

\begin{enumerate}
	\item Consider the system of PDEs of the form 
	\begin{equation}
	F_{m}(\phi_{x},\phi_{t},\phi_{xx},\phi_{tt},\ldots)=0. \tag{B.1}\label{B.1}
	\end{equation}
	Introduce the variable
	\begin{equation}
	\Xi=\mathbf{x}-wt, \tag{B.2}\label{B.2}
	\end{equation}
	where $w<\infty$ is the velocity of the wave profile. Therefore, the system in \myeqref[Equation~]{B.1} becomes
	\begin{equation}
	G_{m}\left(U,\frac{dU}{d\Xi},\frac{d^{2}U}{d\Xi^{2}},\ldots\right)=0. \tag{B.3}\label{B.3}
	\end{equation}
	\item Consider solutions to \myeqref[Equation~]{B.3} of the form
	\begin{equation}
	U(\Xi)=
	\begin{cases}
	\lambda\cos^{\beta}\left(\chi\Xi\right), \quad\quad \vert\Xi\vert\leq\frac{\pi}{2\chi}, \\
	\lambda\sin^{\beta}\left(\chi\Xi\right), \quad\quad \vert\Xi\vert\leq\frac{\pi}{\chi},
	\end{cases} \tag{B.4}\label{B.4}
	\end{equation}
	where $\lambda, \beta$, and $\chi$ are constants to be determined.\\
	\item Take the associated derivatives of \myeqref[Equation~]{B.4} and substitute them into the system of ODEs in  \myeqref[Equation~]{B.3}, while making use of the trigonometric identity
	\begin{equation}
	\sin^{2}\left(\chi\Xi\right)+\cos^{2}\left(\chi\Xi\right)=1. \tag{B.5}\label{B.5}
	\end{equation}
	\item Group the resulting expressions according to the powers of $\sin\left(\chi\Xi\right)$ or $\cos\left(\chi\Xi\right)$. Thereafter, balance the powers, and equate the coefficients to obtain a system of nonlinear algebraic equations. \\
	\item Solve the resulting system of equations, and re-introduce the original variables in order to attain the solution. 
\end{enumerate}

\subsection{The Multiplier / Direct Method}\label{TMDM}

Firstly, for a system of $\mathfrak{m}$ independent variables, we note the definition of the Euler-Lagrange derivative

\begin{equation}
\frac{\delta}{\delta u^{\alpha}}=\frac{\partial}{\partial u^{\alpha}}+\sum_{r\geq 1}D_{i_{1}}, D_{i_{2}},\ldots,D_{i_{r}}\frac{\partial}{\partial u^{\alpha}_{i_{1}i_{2}\ldots i_{r}}},\quad\quad \alpha=1,2,\dots,\mathfrak{m}. \tag{C.1}\label{C.1}
\end{equation}

This method was proposed by Anco and Bluman \cite{AncoBluman1, AncoBluman2, AncoBluman3} as a generalization of Noether's theorem. We algorithmically outline the application of the method below.

\begin{enumerate}
	\item For a given system of PDEs, $F_{m}\left(\mathbf{x},u,\partial u, \partial^{2}u,\ldots,\partial^{k}u\right)=0$, where 
	\begin{equation}
	\partial^{p}=\left\{\frac{\partial^{p}u^{\mu}(\mathbf{x})}{\partial x_{i_{1}}\partial x_{i_{2}}\ldots\partial_{i_{p}}}\;\bigg|\;\mu=1,2,\ldots, n; i_{1},i_{2}\ldots, i_{p}=1,2,\ldots, q\right\}, \tag{C.2.1}\label{C.2.1}
	\end{equation}
	we seek multipliers of the form
	\begin{equation}
	\left\{\Lambda_{m}\left(\mathbf{x},u,\partial u,\partial^{2}u,\ldots,\partial^{\ell}u\right)\right\}_{m=1}^{q}, \tag{C.2.2}\label{C.2.2}
	\end{equation}
	for some specified order, $\ell$. It is imperative that we choose the dependence of the multipliers in a manner such that no singular multipliers arise. \\
	\item Apply the Euler-Lagrange derivative in \myeqref[Equation~]{C.1} to the product of the multiplier and the system of PDEs in order to obtain a set of determining equations. Thereafter, solve the set of determining equations so as to determine all the multipliers. \\
	\item Find the corresponding fluxes that satisfy the identity 
	\begin{equation}
	\Lambda_{m}\left(\mathbf{x},u,\partial u,\partial^{2}u,\ldots,\partial^{\ell}u\right)F_{m}\left(\mathbf{x},u,\partial u,\partial^{2}u,\ldots,\partial^{k}u\right)=D_{i}\varphi^{i}. \tag{C.3}\label{C.3}
	\end{equation}
	\item Each set of fluxes and multipliers yield a local conservation law of the form
	\begin{equation}
	D_{i}\varphi^{i}\left(\mathbf{x},u,\partial u,\partial^{2}u,\ldots, \partial^{r}u\right)=0. \tag{C.4}\label{C.4}
	\end{equation}
\end{enumerate}

\subsection{Lie Symmetries}\label{LS}

Following the notation of Olver \cite{Olver}, consider the $n^{\text{th}}$-order PDE

\begin{equation}
\Theta(x,t,\phi,\phi_{x},\phi_{t},\phi_{xx},\phi_{tt},\dots)=0, \tag{D.1}\label{D.1}
\end{equation}

where $\phi=\phi(x,t)$. \myeqref[Equation~]{D.1} admits Lie point symmetries of the form

\begin{equation}
\Gamma=\xi(x,t,\phi)\frac{\partial}{\partial x}+\tau(x,t,\phi)\frac{\partial}{\partial t}+\eta(x,t,\phi)\frac{\partial}{\partial\phi}. \tag{D.2}\label{D.2}
\end{equation}

The $m^{\text{th}}$-order prolongation is given by
\begin{align}
\text{pr}^{\left[2\right]}\Gamma=&\;\Gamma^{\left[2\right]}=\Gamma+\zeta_{x}\frac{\partial}{\partial\phi_{x}}+\zeta_{t}\frac{\partial}{\partial\phi_{t}}+\zeta_{xx}\frac{\partial}{\partial\phi_{xx}}+\zeta_{tt}\frac{\partial}{\partial\phi_{tt}}, \tag{D.3.1}\label{D.3.1} \\
\zeta_{x}=&\;D_{x}\left(\eta\right)-D_{x}\left(\xi\right)\phi_{x}-D_{x}\left(\tau\right)\phi_{t}, \tag{D.3.2}\label{D.3.2} \\
\zeta_{t}=&\;D_{t}\left(\eta\right)-D_{t}\left(\xi\right)\phi_{x}-D_{t}\left(\tau\right)\phi_{t}, \tag{D.3.3}\label{D.3.3} \\
\zeta_{xx}=&\;D_{x}\left(\zeta_{x}\right)-D_{x}\left(\zeta_{x}\right)\phi_{xx}-D_{x}\left(\tau\right)\phi_{tx}, \tag{D.3.4}\label{D.3.4} \\
\zeta_{tt}=&\;D_{t}\left(\zeta_{t}\right)-D_{t}\left(\xi\right)\phi_{xt}-D_{t}\left(\tau\right)\phi_{tt}, \tag{D.3.5}\label{D.3.5}
\end{align}

and $D_{i}$, for $i\in\left\{x,t\right\}$, is the operator of total differentiation defined as

\begin{equation}
D_{i}=\frac{\partial}{\partial x^{i}}+u^{\alpha}_{i}\frac{\partial}{\partial u^{\alpha}}+u^{\alpha}_{ij}\frac{\partial}{\partial u^{\alpha}_{j}}+\ldots+u^{\alpha}_{ii_{1}i_{2}\ldots i_{\ell}}\frac{\partial}{\partial u^{\alpha}_{i_{1}i_{2}\ldots i_{\ell}}}+\ldots, \quad\quad i=1, 2, \ldots, \mathfrak{n}. \tag{D.4}\label{D.4}
\end{equation}

The application of \myeqref[Equation~]{D.3.1} on \myeqref[Equation~]{D.1} yield a system of linear PDEs

\begin{equation}
\xi\frac{\partial\Theta}{\partial x}+\tau\frac{\partial\Theta}{\partial t}+\eta\frac{\partial\Theta}{\partial\phi}+\zeta_{x}\frac{\partial\Theta}{\partial\phi_{x}}+\zeta_{t}\frac{\partial\Theta}{\partial\phi_{t}}+\zeta_{xx}\frac{\partial\Theta}{\partial\phi_{xx}}+\zeta_{tt}\frac{\partial\Theta}{\partial\phi_{tt}}=0. \tag{D.5}\label{D.5}
\end{equation} 

Equating the various powers of the variable, $\phi$, in \myeqref[Equation~]{D.5} to zero produces a set of linear PDEs. Solving the system of PDEs provide a functional form for $\xi, \tau,$ and $\eta$, which generate each Lie point symmetry.  

\section*{Acknowledgements}
\noindent MAZK, MM, and FP thank the University of KwaZulu-Natal of the Republic of South
Africa for making this research possible. FP acknowledges that this research is supported
by the South African Research Chair Initiative of the Department of Science and Innovation
and the National Research Foundation (NRF). \\


The authors declare that they have no conflict of interest.

\end{document}